\pdfoutput=1
\documentclass[11pt,a4paper]{article}
\usepackage{array}
\usepackage[footnotesize,bf]{caption}
\usepackage[dvips]{graphicx}
\usepackage{color} 
\usepackage{amsmath}
\usepackage{fancyhdr}
\usepackage{natbib}
\usepackage[LGRgreek]{mathastext}
\usepackage[yyyymmdd,hhmmss]{datetime}

\fancyheadoffset[R]{0pt}
\fancyfootoffset[R]{0pt}
\renewcommand{\headrulewidth}{0.4pt}
\renewcommand{\footrulewidth}{0.4pt}

\definecolor{orange}{rgb}{1.0,0.5,0.0}
\definecolor{aqgr}  {rgb}{0.0,1.0,0.6} 
\definecolor{viol}  {rgb}{0.8,0.6,0.8}
\definecolor{figdr} {rgb}{1.0,1.0,1.0} 
\definecolor{colnu} {rgb}{1.0,0.0,1.0} 
\definecolor{colhd} {rgb}{1.0,0.8,0.0} 

\newcolumntype{C}[1]{>{\centering\let\newline\\\arraybackslash\hspace{0pt}}m{#1}}
\title{\bfseries{\textsc{Quadripolar Relational Model: \\
   a framework for the description of borderline \\
   and narcissistic personality disorders}}} \date{}
   
\author{Alessandro Fontana}

\begin{document}
\maketitle
   
\clubpenalty=10000
\widowpenalty=10000

\begin{abstract}
Borderline personality disorder and narcissistic personality disorder are important nosographic entities and have been subject of intensive investigations. The currently prevailing psychodynamic theory for mental disorders is based on the repertoire of defense mechanisms employed. Another line of research is concerned with the study of psychological traumas and dissociation as a defensive response. Both theories can be used to shed light on some aspects of pathological mental functioning, and have many points of contact. This work merges these two psychological theories, and builds a model of mental function in a relational context called Quadripolar Relational Model. The model, which is enriched with ideas borrowed from the field of computer science, leads to a new therapeutic proposal for psychological traumas and personality disorders.
\end{abstract}


\section{Introduction}  

This work presents a model of emotional interaction, called \textbf{Quadripolar Relational Model}. Most models of emotions emerged in the last decade \citep{marsella2010} belong to one of the following categories: anatomic models, dimensional models, appraisal models. Anatomical models \citep{ledoux2000} try to reconstruct the neural pathways that underlie emotional reactions; one of the best known results is the elucidation of the role of amygdala in fear processing. In dimensional models \citep{rubin2009} emotions are characterized by a number of axes or dimensions, and the emotional state of an individual is represented as a point of a hypercube.  

In appraisal models, emotions are argued to arise from patterns of individual judgment concerning the relationship between events and personal beliefs, desires and intentions \citep{lazarus1991, sander2005}. For example, if external reality is congruent with the person's life goals, he/she may feel happy. Appraisals may also trigger cognitive responses, often referred to as coping strategies (e.g., planning, procrastination or resignation) feeding back into a continual cycle of appraisal and re-appraisal.

As we shall see, our model foresees the existence of four relational states, each associated to specific emotions. The concept of relational states is present in transactional analysis \citep{berne1964}, a psychological theory which considers the self as consisting of three structures or states, each with specific functions: the parent, the adult and the child. The communication between two individuals can be read as a ``transaction'' (or exchange) between different states or counterparts of two selves. 

Another relevant model is Structural Analysis of Social Behavior (SASB) \citep{benjamin1996}, which offers a framework to describe human relationships as points in a three-dimensional space. The three axes are: ``love-hate'', ``enmeshment-differentiation`` and ``interpersonal focus''. Points between poles represent behavioral patterns (such as ``blame'', ``attack'' and ``protect'') consisting of components from the nearest poles. The central idea of SASB-based therapy, which was applied to personality disorders, is that patterns of disorder are replications of patterns learned from caregivers. 

Our model merges psychodynamic elements with elements borrowed from the theory of psychological traumas and builds a framework for the description of borderline personality disorder and narcissistic personality disorder. A thorough characterization of these complex conditions, which would require a detailed description of symptoms and a consideration of various sub-types, is outside the scope of this work: our objective is to put in evidence some key features, that we will then try to model. We are aware of the complexity of psychological disorders: our model, based on a simplification of reality, has the purpose to interpret some limited aspects of mental functioning.

\textbf{Borderline personality disorder (BPD)} \citep{gunderson2005} is characterized by a pervasive instability of mood, of personal relationships, of the perception of the self, identity and behavior. The central defense mechanism used by persons with BPD is splitting, which can be defined as the inability to integrate positive and negative aspects of the self and others. The result is a characteristic view of the world in ``black and white'' \citep{perry2013}. Another typical defense mechanism employed in BPD is projective identification, by which split-off parts of the self are projected into the object to exert possess and control \citep{klein1946}. As a result of using such defenses, persons with BPD often engage in idealization and devaluation of others, alternating between high positive regard and heavy disappointment.

The main characteristics of \textbf{narcissistic personality disorder (NPD)} \citep{lowen2004} are an excessive reference to others for self-esteem regulation, a deficit in the ability to feel empathy towards other individuals, a particular self-perception of the subject defined ``grandiose Self'', and a difficulty of emotional involvement. The exaggerated dependence on others for approval causes self-appraisal to oscillate between extremes. The person is characterized by a form of deep selfishness of which he/she is usually not aware. The consequences are likely to produce suffering, social problems or significant emotional and relational difficulties. NPD shares with BPD the use of splitting and projective identification as core defense mechanisms. 

The currently prevailing psychodynamic vision \citep{ahles2004} characterizes mental diseases (including BPD and NPD) based on the repertoire of associated defense mechanisms. Such defenses would be erected to overcome intrapsychic conflicts and/or wounds originating in the interaction between the future patient and the caregiver. According to O. Kernberg BPD and NPD share a so-called ``borderline personality organization'' \citep{kernberg1967}, conceptualized in terms of unintegrated and undifferentiated affects and representations of the self and others.

The ``psychotraumatologic vision'', instead, is concerned with the investigation of psychological traumas of different kinds and the associated mental structures. The most common response to trauma is represented by dissociation, defined as the distortion, limitation or loss of the normal associative links between perceptions, emotions, thoughts and behavior. The outcome of traumatic experiences is a characteristic syndrome called ``post-traumatic stress disorder'' (PTSD) \citep{andreasen2010ptsd}, which can be ``complex'' when the traumas are repeated, often within a familiar context \citep{herman1992complex}.

In both BPD and NPD the unstable sense of self can lead to episodes of dissociation and chronic feelings of emptiness. Dissociation can take the form of depersonalization (feeling of separation from oneself and one's body), derealization (feeling of being detached from the external world), selective amnesia and emotional detachment \citep{radovic2002depers}.

The rest of the paper is organized as follows. The model is introduced in section 2 and used in section 3 to obtain an interpretation of some aspects of BPD and NPD. Section 4 puts forward a therapeutic proposal based on the model. Section 5 draws the conclusions and outlines future research directions.

\section{The quadripolar relational model}

\subsection*{Value of features}  

External reality can be conceptualized in terms of \textbf{features}. Examples of simple perceptual features are: shape, orientation, color. Examples of more abstract features are: ``black horses'', ``boys playing tennis'', ``to be a teacher''. Features are not independent of each other: complex features (e.g. faces) are based on simpler ones (e.g. oriented edges). Features can be conceived as nodes of a (very large) neural network in the brain.

Features can be active or inactive and different situations correspond to different sets of active features. For instance, if a person is walking in the jungle, the features encoding the concepts of ``dark'' and ``tree'' may be active, while other features, such as ``office chair'' and ``computer screen'', are likely to be inactive. Feature activation and deactivation is a continuous process, driven by perceptual features fed from external stimuli, and propagated to more abstract ones in real time. This occurs on a fast time scale, as the mind ``navigates'' through everyday life. 

We postulate that features are characterized by a property called \textbf{value}, that can be either positive or negative. Features such as ``beauty'' or ``health'', for example, may be considered positive, while features such as ``deception'' or ``illness'' are generally considered negative. The determination and change of feature value happens by association: if a new feature of unknown value is presented in association with positive features, it will assume a positive connotation. Since a feature in general takes part in many associations, its value will be determined by the combined effect of all associations. Feature value is a long term property, expected to change on a slower time scale. 

\subsection*{Relational module}  

Human beings are social animals and relations play a central role in human behavior. Emotions arise often as a reaction to the actions of another human being within a relational context. Humans give also much importance to the role of hierarchy. In many life situations, it is crucial to decide who is right and who is wrong, who is superior and who is inferior. This can be helpful when an employee must decide which tone he/she should use while addressing one of his/her peers or a senior board member. In some circumstances this insight can be a matter of life and death.

\begin{figure}[t] \begin{center} \hspace*{-0.0cm}
{\fboxrule=0.0mm\fboxsep=0mm\fbox{\includegraphics[width=16.00cm]{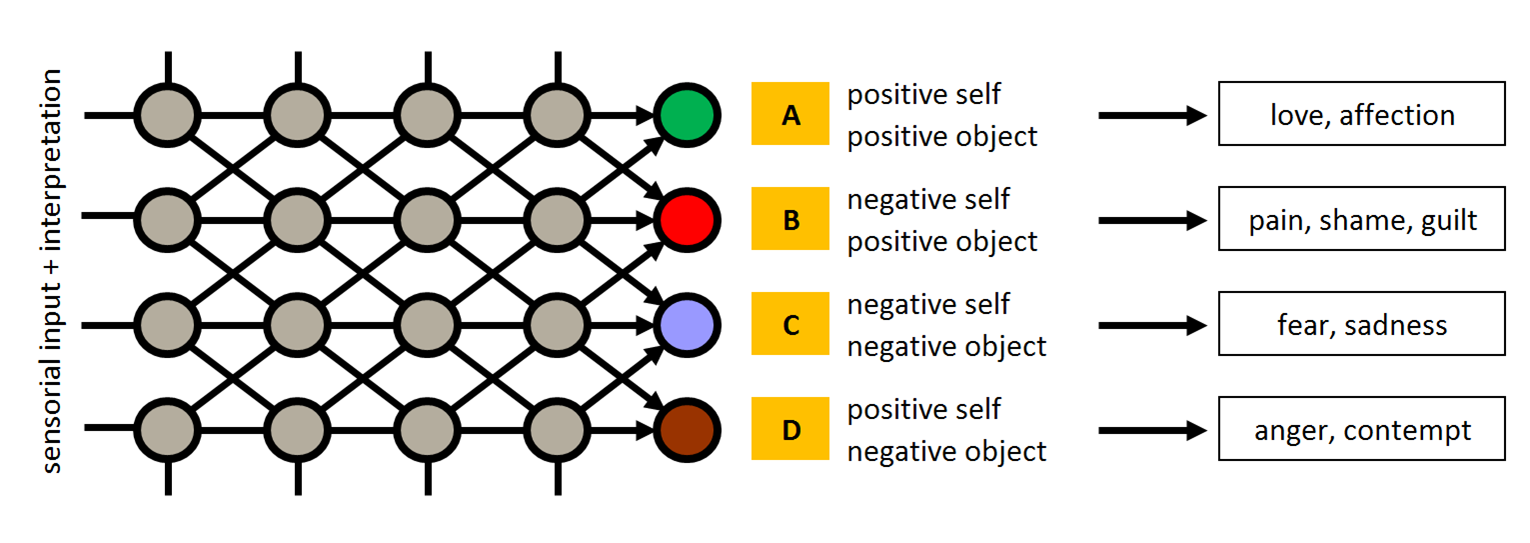}}}
\caption{Neural module that manages relational judgements. The module classifies the sensory input into four poles. Each pole corresponds to a different combination of states for the self and for the object, and is associated to specific emotions.}
\label{relnet}
\end{center} \end{figure}

Our assumption is that a specific neural module is dedicated to managing relational judgements (Fig.~\ref{relnet}). This module makes use of the stimuli coming from the perceptual apparatus, as well as of information obtained from memory, and attributes a value to two key symbols: the self and the object. In this context, ``positive'' can be interpreted as either ``important'' or ``right'', while ``negative'' can be interpreted as either ``unimportant'' or ``wrong''. As a result, there are  four possible relational poles, each associated to specific emotions: 

\begin{itemize}
\item A : positive self and positive object \\ (the self feels love and affection towards a rewarding object)
\item B : negative self and positive object \\ (the self feels pain and shame in front of a superior /humiliating object)
\item C : negative self and negative object \\ (the self feels sadness or fear when self and object are both negative or helpless)
\item D : positive self and negative object \\ (the self feels anger for a guilty object or despises an inferior object)
\end{itemize}

The relational module has the objective to process the perceptual input, produce an interpretation of reality and select one of these relational poles. In this sense, our model may be considered as belonging to the category of appraisal models of emotions. The association of shame, guilt and sadness to a lower social status is confirmed by numerous studies \citep{stevens1996}. Also the association of anger to a higher social status has been recognized and studied \citep{tiedens2011}.


\subsection*{Normal functioning}  

\begin{figure}[t] \begin{center} \hspace*{-0.50cm}
{\fboxrule=0.0mm\fboxsep=0mm\fbox{\includegraphics[width=18.00cm]{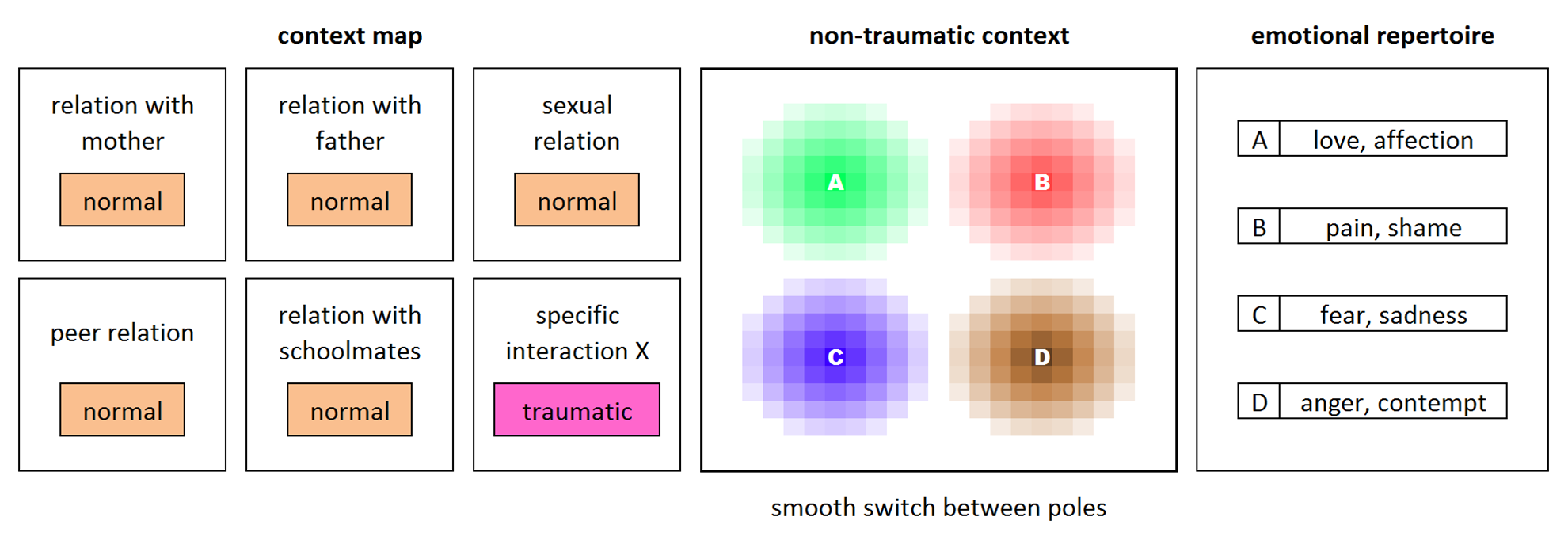}}}
\caption{Normal condition. In the non-pathological case traumatic contexts are very few and limited in scope (left panel). The mind can switch smoothly between relational poles in each context, thanks to an intact relational module (middle panel). No pole requires dissociation: the repertoire of emotions is fully accessible (right panel).}
\label{switchx}
\end{center} \end{figure}

From a relational perspective, the objective of mental activity is to create a map of all possible situations that can be encountered in the course of life, and build for each situation a model that tells how to behave. Relational situations can be grouped in different ``contexts'', such as ``relation with mother'', ``peer relation'',``relation with schoolmates'' (Fig.~\ref{switchx}-left). Each context contains situations that are similar and require a specific approach: when the same or a similar situation is encountered again, the relevant model is retrieved from memory and used to direct behavior.

A context can be represented on a plane, in which points correspond to individual situations (Fig.~\ref{switchx}-center), characterized by different combinations of active features. Each point belongs to the ``zone of influence'' of one of the four relational poles. Each pole originates from a point representing the most prototypical situation associated to the pole, and extends towards less prototypical situations. Pole D, for instance, may originate from a situation-point characterized by an object behaving very dishonestly, eliciting a very strong anger. Points further away from the center of the pole may be characterized by a better behavior of the object. In normal conditions, the mind can switch smoothly between poles, during its navigation through everyday life.

\subsection*{Traumatic events}

The relational poles that can be involved in a trauma are those indicated with B and C in Fig.~\ref{relnet}, in which the self is negative: we refer to them as to the \textbf{negative poles}, for brevity. In case they are affected by a trauma, they will be called \textbf{traumatic poles}. Pole B consists of a negative self and a positive object. This may correspond to a situation in which a child is abused physically or verbally by the caregiver. Pole C corresponds to a self and an object which are both negative and subject to intense fear and/or sadness and/or helplessness. This might occur when the caregiver is unable to calm the child, and actually contributes to increase his/her fear showing a frightened expression. This is the so-called frightened-frightening caregiver which is associated to forms of disorganized attachment \citep{main1986}.

In case of trauma, the emotional values elicited are too high and the functioning of the relational module becomes critical, not unlike what happens to a camera when the light is too intense and the picture becomes blurred. We assume that the level of emotion E is proportional to the difference between the perceived value of the object $Vobj$ and the perceived value of the self $Vself$ in case B, and to the difference between a reference value $Vref$ and the perceived value of the self in case C. The value of the self and of the object are in turn determined by the sum of the values of all active features associated to them, $Vs_{i}$ and $Vo_{i}$. In formulas (S is the sigmoid function, which prevents the emotional values from exceeding a maximum threshold):

\begin{flushleft}
Pole B: $ E = S(Vobj-Vself) $ \\
Pole C: $ E = S(Vref-Vself) $ \\ 
$ Vself = S(\sum_{i}Vs_{i}); Vobj = S(\sum_{i}Vo_{i}) $ \\
\end{flushleft}

An example can clarify how the occurrence of a trauma is modeled. Let us suppose that the victim of the trauma is a violin player and the perpetrator is the orchestra conductor. Before the trauma, in the victim's mind there are features associated to the self: ``not being an artist'' (negative), ``imprecision'' (negative), ``to make mistakes while playing the violin'' (negative). In the victim's mind, there are also features associated to the object: ``to be a great artist'' (positive), ``perfection'' (positive), ``to make no mistakes'' (positive). If, during a concert, the victim makes a mistake and is criticized by the conductor, the value of the self-related features are further reduced and drag down the value of the self, while the value of the object remains very high. The level of pain and shame become extremely high and the trauma takes place. 

This scenario is very likely to occur within a family, where the values of all features are set by the parents. In other words, the parents decide the rules of the game and the child has no choice but to accept to play a game governed by rules on which he/she has no say. This gives an unfair advantage to the parents, who are likely to be better than the child at their own game. This of course does not need to lead to a trauma, if the parents do not critisize or humiliate the child for not ``playing well''.

\begin{figure}[t] \begin{center} \hspace*{-0.0cm}
{\fboxrule=0.0mm\fboxsep=0mm\fbox{\includegraphics[width=17.00cm]{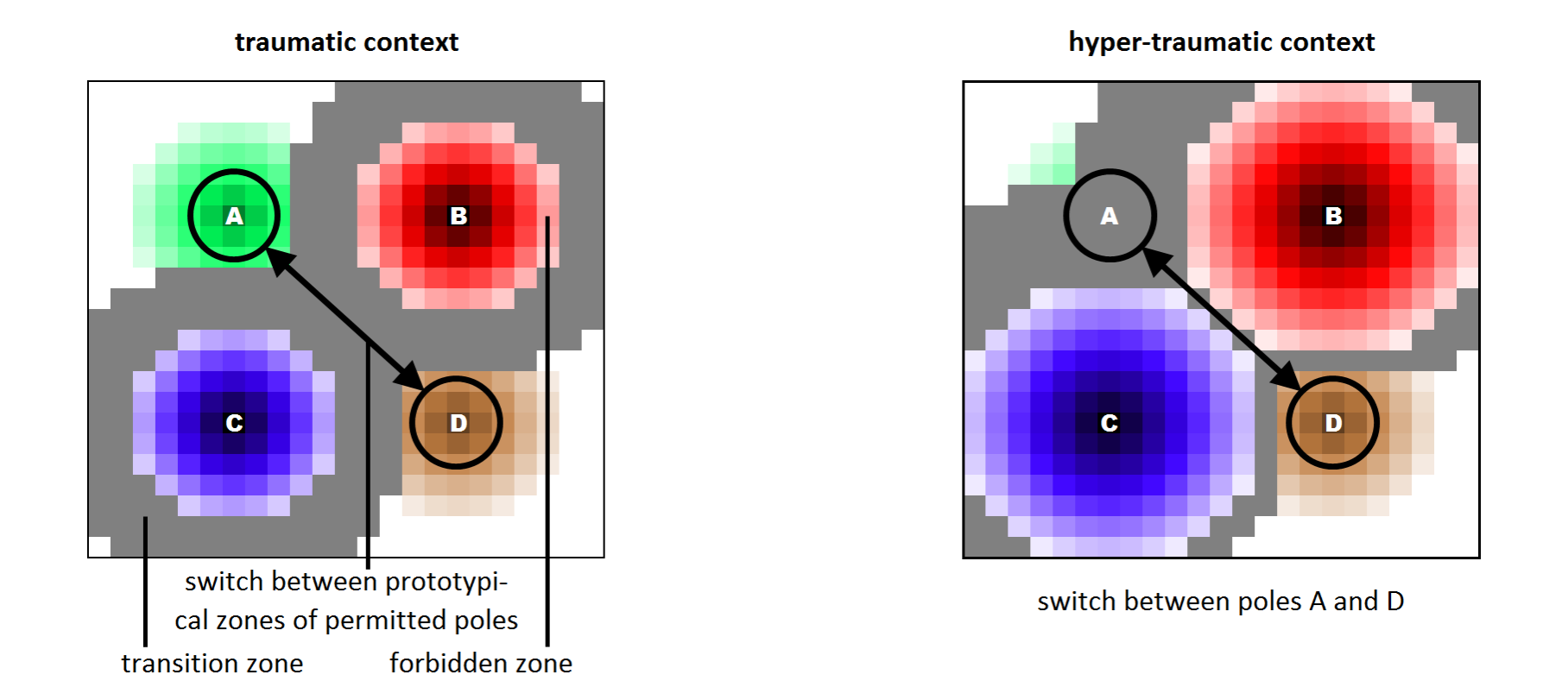}}}
\caption{Map of relational poles in the feature space in case of trauma. In a traumatic context (left), pole B and/or pole C are characterized by too high emotional levels, and are inhibited. When the mind happens to be in one of these poles (forbidden zone), emotional dissociation kicks in. To avoid dissociation, the mind oscillates between poles A and D. The mind can stay in pole A or pole D as long as the situation is prototypical for the pole. When the situation drifts out of the prototypical zone into the transition zone (shown in grey), splitting symptoms appear, and the mind switches to the prototypical zone of the other permitted pole. For a hyper-traumatic context (right), the forbidden and transition zones are much more extended and the free zone around pole A vanishes.}
\label{bubbles}
\end{center} \end{figure}

\subsection*{Response to trauma: dissociation}
Upon the occurrence of a trauma, the associative function of the mind stops working: this corresponds to the phenomenon of dissociation. We distinguish between two different concepts of dissociation. The first concept, which we call ``online dissociation'', refers to the selective exclusion from consciousness of one or more perceptual channels. This is a temporary condition, caused by the fact that some emotional values are too high. The second concept is that of ``memory dissociation'', or ``dissociated memory''. This concept refers to a condition of the memory, in which there are incomplete records, that for this reason exhibit a lower degree of connectivity with the rest of memory and are more difficult to retrieve. A dissociated memory is the outcome of episodes of online dissociation.  

When can hypothesize that, upon the first occurrence of a trauma, dissociation is deep and involve both the rational and the emotional part. This may correspond to the ``freezing'' behavior of children while interacting with their mothers, in case of ``disorganized attachment'' \citep{main1986}. This kind of dissociation provides an effective shield to emotional pain. On the other hand, a condition in which mental function is temporarily turned off exposes the individual to serious consequences in a potentially dangerous environment. The use of deep dissociation as a defense mechanism is impractical, potentially fatal, and cannot be sustained for long periods of time. 

Emotional dissociation is a more limited form of dissociation which, in a traumatic situation, excludes from consciousness the emotional component, while the rational part continues to be active. The subjective experience that accompanies this kind of dissociation may be represented by the phenomena of derealization and depersonalization, characterized by a diminished sense of reality. It is not unrealistic to assume that, in case of repeated traumas, the mind learns to replace total dissociation with emotional dissociation, as a compromise able to allow a higher level of functioning.  


\subsection*{Response to trauma: splitting}

In case of trauma, the adoption of emotional dissociation makes it possible for the mind the stay in the negative relational poles B and C, excluding intolerable emotions from consciousness. It is compatible with a mental functioning which would appear normal to most observers. However, it implies the absence of the emotional component of mental life, which represents a heavy price to be paid. Therefore, the mind tries to avoid the negative poles B and C, and heads towards the positive poles A and D, where it can exist with all its functions active: this corresponds to the defence mechanism of splitting.

In the presence of traumas, the context space can be divided into three zones (Fig.~\ref{bubbles}-left): ``forbidden zone'', ``transition zone'' and ``free zone''. The forbidden zone is the zone around a traumatic pole, which cannot be accessed in a non-dissociated state (hence the term ``forbidden''); the transition zone is a safety belt around the forbidden zone, and the free zone is an area far from all traumatic poles, corresponding to the zone around the positive poles A and D. 

The mind stays in the zone around pole A as long as conditions are prototypical for pole A, i.e. as long as the relation with the object is perfect, full of trust and mutual respect. As conditions depart from their prototypical values for pole A, the mind switches abruptly to the zone around pole D. When conditions are not perceived as prototypical for pole D, the mind jumps back to pole A, and the cycle repeats itself. This model explains the idealization-devaluation cycle based on splitting. Fig.~\ref{bubbles}-right shows a context in which the impact of trauma is deeper and, as a result, the forbidden and transition zones are larger: we call such context ``hyper-traumatic''.

The transition zone is a safety belt built around the forbidden zone. In the transition zone the features involved in the trauma (the defects criticized) appear distorted and magnified. The distortion of defects serves the purpose to alert that the current situation is not prototypically positive anymore and it is creeping towards a traumatic pole. With reference to the example of the violin player, the slightest mistake done when playing in public may sound magnified and ``threatening'' in the player's mind. If the feature criticized is ``to have a big nose'', when looking at the mirror the person will see his/her nose magnified, like in a caricature. Sometimes, despite all efforts, the mind cannot avoid to fall back into a traumatic pole. When this happens, it resorts to emotional dissociation to protect itself from pain.


\section{Interpretation of personality disorders}

\subsection*{Interpretation of BPD}

\begin{figure}[t] \begin{center} \hspace*{-0.50cm}
{\fboxrule=0.0mm\fboxsep=0mm\fbox{\includegraphics[width=18.00cm]{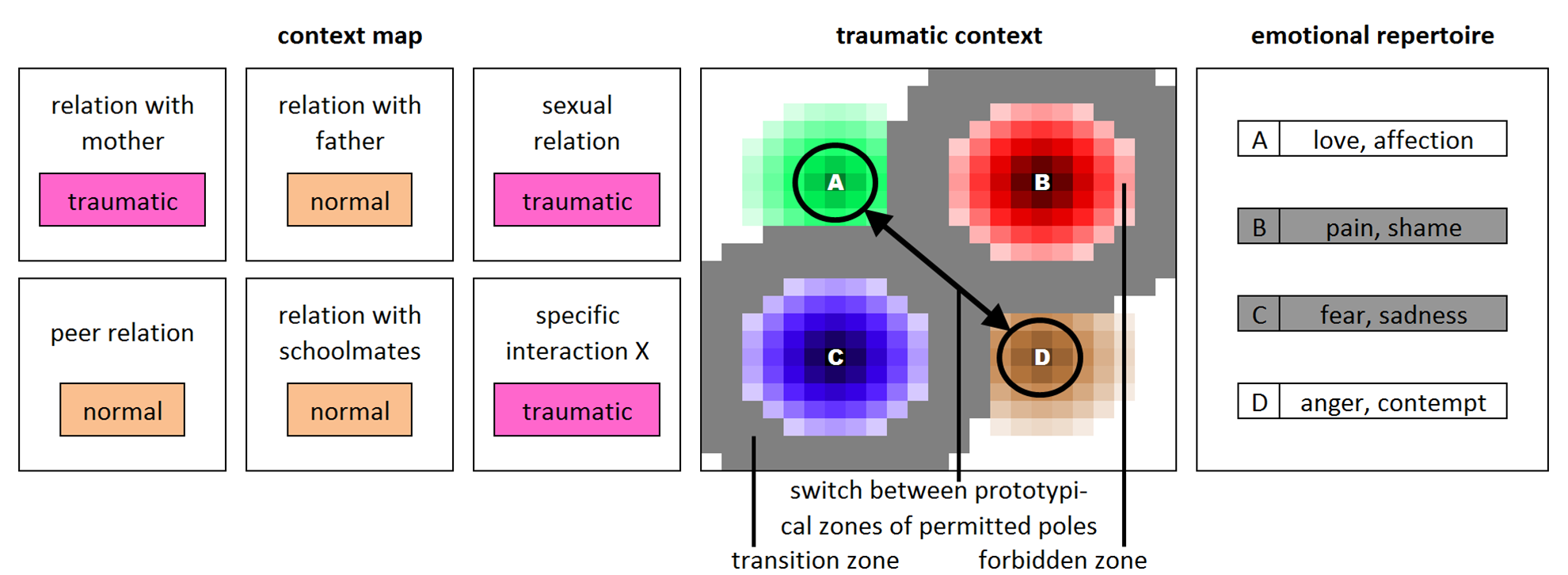}}}
\caption{BPD: context map and emotional functioning. Some key contexts, such as ``mother relation'' and ``sexual relation'', are traumatic. In these contexts the functioning of the relational module is impaired: poles B and C are difficult to access in a non-dissociated condition, and the mind oscillates between poles A and D.}
\label{switchb}
\end{center} \end{figure}

BDP is a complex condition, in which genetic and environmental factors are hypothesized to play a role. Childhood traumas involving caregivers are often present in the history of persons with BPD \citep{macintosh2015, elices2015}. As a result, the functioning of the relational module may be impaired in key contexts such as ``mother relation'' and ``sexual relation'' (Fig.~\ref{switchb}). In such contexts poles B and C are difficult to access, and the mind is forced to oscillate between poles A and D. When the mind falls into poles B or C, emotional dissociation is used to avoid re-living the painful emotions associated to the trauma.

Despite the evident drawbacks of this behavior, the BPD person manages to keep his/her emotions while oscillating between the two sides of splitting, and only occasionally resorts to emotional dissociation, as a last defense line. The maintenance of the emotional part allows the BPD person to spend most of the time on a relatively high level of functioning. The BPD person is not happy when he/she experiences emotional detachment, he/she understands that something is wrong.

In the transition zone, the distortion and amplification of the features involved in traumatic situations has the purpose to warn the mind that it is nearing a traumatic pole. This is reminiscent of the phenomenon of ``aberrant salience'' which, according to some theories \citep{kapur2003salience}, is deemed responsible for the positive symptoms of schizophrenia (hallucinations and delusions). The salience mechanism would be mediated by the neurotransmitter dopamine: this would explain why antipsychotic drugs, that act by blocking the dopamine receptor, are effective in eliminating hallucinations. In contrast to the ``pure'' dopamine hypothesis of schizophrenia \citep{baumeister2002}, these theories explain why antipsychotic drugs have effect only after a period of time after the start of their administration. 

An intriguing possibility suggested by our model is that the phenomenon of aberrant salience in schizophrenia and the mechanism of feature distortion in the transition zone of a traumatic context could both be expressions of a standard warning mechanism used by the mind to stay away from traumatic poles. Transitory perceptual distortions or ``pseudo-hallucinations'' are present also in personality disorders \citep{gras2014}. The difference between such distortions and schizophrenic hallucinations would be quantitative rather than qualitative. The higher degree of severity of the hallucinatory symptoms in schizophrenia could be determined by a heavier traumatic charge and possibly mediated by a genetic predisposition. Another clinical example that could be explained by the phenomenon of feature distortion is the dysmorphic body image associated to eating disorders (among many other disorders) \citep{ruffolo2006}. 

\subsection*{Interpretation of NPD}

\begin{figure}[t] \begin{center} \hspace*{-0.50cm}
{\fboxrule=0.0mm\fboxsep=0mm\fbox{\includegraphics[width=18.00cm]{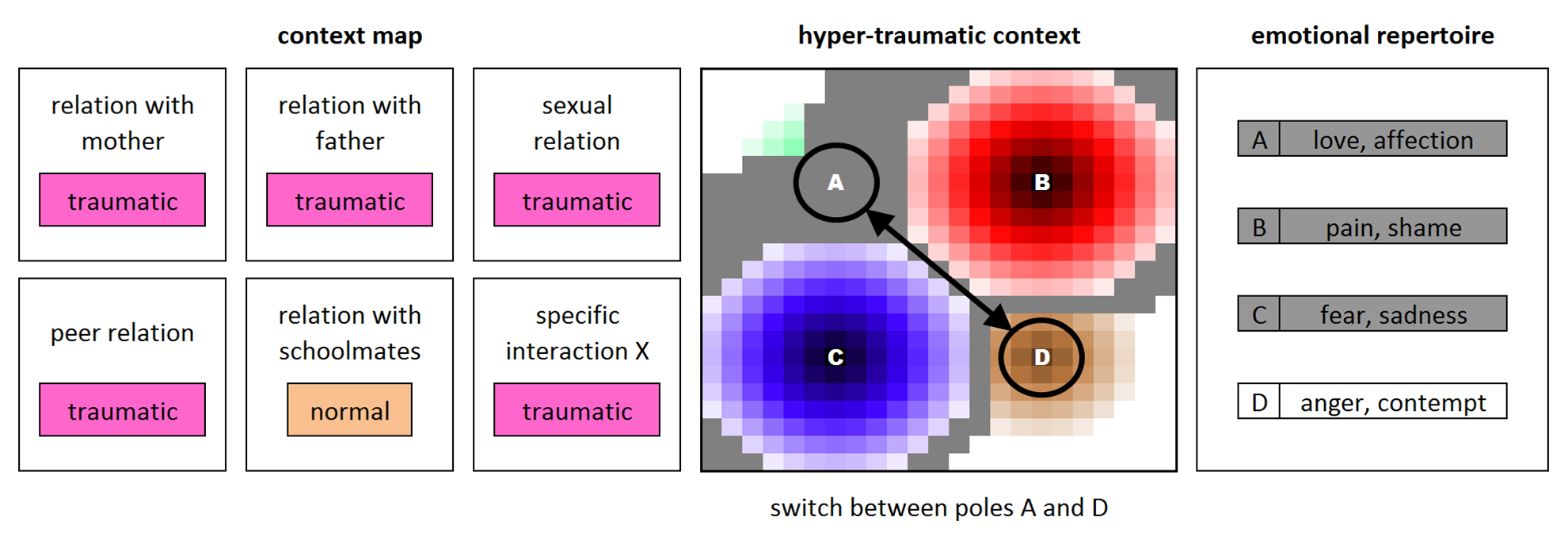}}}
\caption{NPD: context map and emotional functioning. Some key contexts, such as ``mother relation'' and ``sexual relation'', are traumatic. In these contexts the functioning of the relational module is impaired: poles B and C cannot be accessed without dissociation, and the mind is forced to oscillate between poles A and D. Unlike the BPD case, pole A is devoid of emotional content, and filled with grandiose fantasies.}
\label{switchn}
\end{center} \end{figure}

As in the case of BDP, it is not unreasonable to think that, besides genetic predisposing factors, childhood traumas may play a role in the development of NPD \citep{lowen2004}. As a result, the functioning of the relational module may be impaired in key contexts such as ``mother relation'' and ``sexual relation'' (Fig.~\ref{switchn}). In such contexts poles B and C cannot be accessed without dissociation and the mind is forced to oscillate between poles A and D. When the mind falls into poles B or C, emotional dissociation is used to avoid feeling the painful traumatic emotions.

Our assumption is that, in the NPD case, the traumas affecting a traumatic context are more severe. As a result, the forbidden and transition zones are more extended, so that the free zone around pole A is almost non-existent, while the free zone around pole D may still be present. Splitting is employed also by the NPD person, but with a different connotation. Pole D, in which a positive self unleashes its anger towards a frustrating and guilty object, is present also in NPD: the difference lies in pole A. While in the BPD case pole A is characterized by love and affection, in the NPD case love and affection do not belong to the repertoire of memories of the person, and therefore cannot be used to build a behavioral scheme for pole A. Instead of a scheme based on real life examples, the NPD person resorts to a fantasy: the bright side of his/her splitted reality is played by the grandiose self.

The grandiose image is not associated to strong positive emotions: rather, it is imbued with a sense of victory and revenge for the past frustrations. Since pole A of NPD is not grounded on concrete experiences, is not associated to strong feelings and overall lays on very shaky grounds (such as the reward coming from the external world), it is much more volatile and unstable compared to pole A of BPD. When the world does not show enough empathy and/or provide sufficient reward, the NPD person jumps into pole D or is forced back into emotional dissociation. This is the state in which the NPD person spends most of his/her time. 

This condition, characterized by emotional numbness, is not so disturbing for the NPD person as it is for the BPD person. This because the BPD person lives for most of the time in the world of emotions, and immediately perceives when dissociation kicks-in. The NPD person, on the contrary, is used to living without emotions: this is his/her normal condition. From the viewpoint of the emotional content, switching from the grandiose self to dissociation does not make much of a difference.    

A reduced capacity to feel (certain) emotions is considered a central feature of antisocial personality disorder (APD) and psychopathy \citep{stevens2001facial, hastings2008}. This trait can be interpreted with our model if we assume that, in case of psychopaths, the negative poles B and C are not associated to strong emotional responses and therefore cannot be traumatized. Or, for this category of people, the capacity to dissociate would be so deep and pervasive as to eliminate the need for a transition zone in the feature space. If this were true, APD and psychopathy could be inserted into our framework as extreme cases of NPD. 


\section{Therapeutic proposal}

So far, we have seen how our model requires that psychological traumas that disrupt the functioning of the relational module in pole B are due to an emotional abuse by an object perceived as worthier than the self. In other words, a value gap between the object and the self is a precondition for the establishment and for the maintenance of a trauma.

A line of thought maintains that, to resolve a trauma, it is necessary to relive the traumatic event and reexperience the negative emotions associated, while providing at the same time a more favorable interpretation of the event. This can be done with the help of EMDR \citep{shapiro2010}, which facilitates the reprocessing of the traumatic material. However, if the associated pain is too strong, the patient may not be willing to relive it and, once brought (by the psychologist) to the traumatic point, he/she resort to emotional dissociation. In this way, the trauma is not relived at the emotional level, and is not resolved. We argue that, to ensure that the patient is able to access the traumatic experience emotionally, the level of pain must be reduced \textit{preventively}.

\begin{figure}[t] \begin{center} \hspace*{-0.50cm}
{\fboxrule=0.0mm\fboxsep=0mm\fbox{\includegraphics[width=18.00cm]{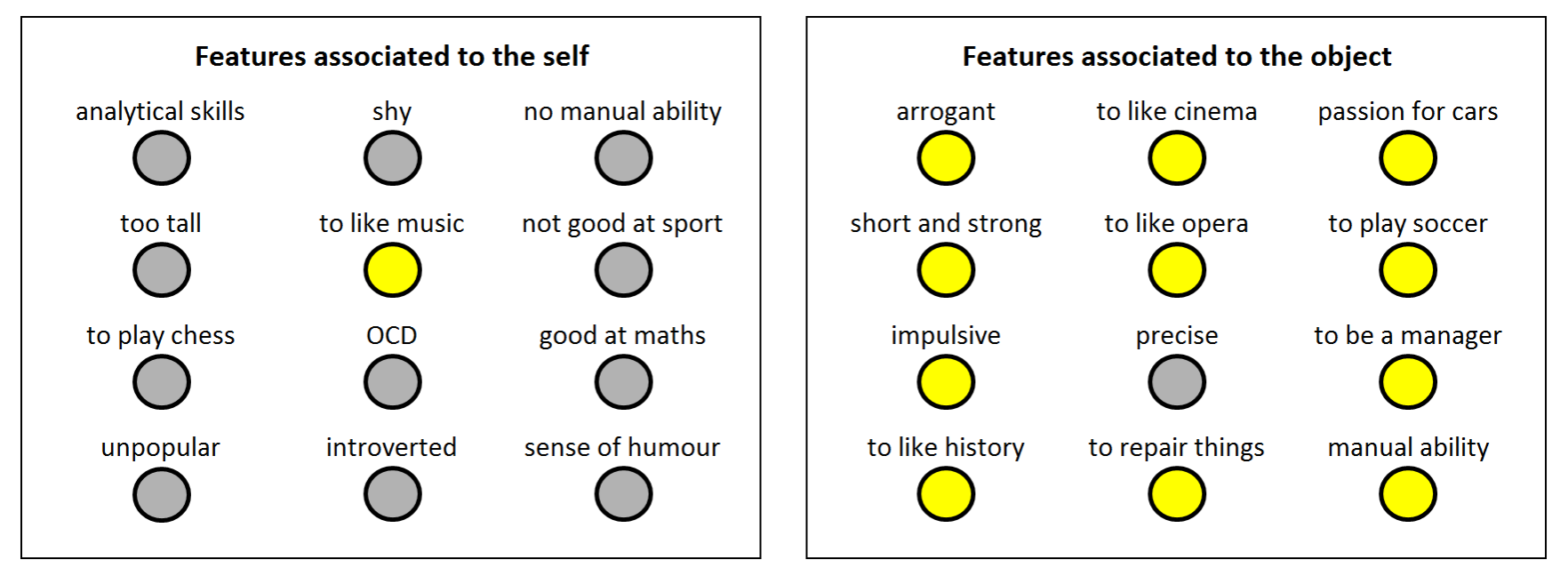}}}
\caption{Features of the self and of the abusing object. Most features associated to the self are negative, most features associated to the object are positive (positive features are shown in yellow, negative features are shown in grey).}
\label{features}
\end{center} \end{figure}

As pointed out in section 2, the level of emotional pain is proportional to the difference between the perceived value of the object and the perceived value of the self. The value of the self and of the object depend in turn on all the associated features. If the value of the self is low, it is because most associated features are negative. If the value of the object is high, it is because most associated features are positive (Fig.~\ref{features}). It is impossible to have a negative self mostly linked to positive features, or a positive object mostly linked to negative features.

In other words, the landscape of associated features constitutes the foundation of the traumatic structure: if the value of such features is changed, we can expect the value of the self and of the object to change accordingly. As the value gap narrows, the level of emotional pain should also diminish. At this point, the trauma would be susceptible of being attacked with techniques such as EMDR, until its full resolution. Therefore, our therapeutic strategy is to address the features associated to the self and to the object and change their value in the patient's mind. 

The good news is that the value attributed to most features is arbitrary. Some features, such as intelligence, beauty, wealth, etc. are universally acknowledged as being positive. For such features we may hypothesize the existence of some kind of genetic-evolutionary basis. But the value of the vast majority of features appears to be completely arbitrary.

This can be appreciated by considering the diversity of convictions, beliefs and reference values across cultures, geographical areas, ethnic groups and historical periods. Being slim nowadays is a very positive feature. This was not the case in the past, when food was in scarse supply and a higher level of fat was linked to a higher social status. Being tanned was once an negative feature typical of members of the working class who had to work in the open, until Coco Chanel convinced us that it was actually a good thing!

A high diversity of patterns of feature values can be seen in the same historical period and within the same nation, across different social networks. For a given family, the feature ``being good at maths'' can be considered positive and be appreciated, and the feature ``play the piano'' can be considered negative and disregarded. In a different family, the opposite may be true. This depends very much on the history and taste of the senior family members (usually the parents), who set the core background value scenario. In principle, nothing prevents to bring changes to this scenario.   

This can be done by means of counterexamples. Given, for instance, the negative feature ``risky'', we can provide counterexamples in which taking risks proved to be a good idea. We might mention Julius Caesar, who chose to cross the Rubicon; Jeff Bezos, who left a well-paid Wall-Street job to create Amazon; Butch Cassidy and Sundance Kid, who jumped into a waterfall (at least in the movie) and saved their lives. Given the negative feature ``unpopular'', we might say that Napoleon was very unpopular at the military academy and Einstein was unpopular at school. Linking negative features to positive examples raises the features' value. Similarly, linking positive features to negative examples can be used to lower the features' value.
  
The therapeutic strategy suggested has some points of contacts with Cognitive Behavioral Therapy (CBT) \citep{longmore2007}. CBT is focused on changing distorted thoughts, dysfunctional emotions and maladaptive behaviors of the patient, in order to reduce the symptoms of a psychological disorder. Unlike other psychotherapies which delve into the patient's past, CBT is mainly concerned with the present and oriented to the specific symptoms.

There is a key difference between our suggested therapeutic framework and CBT. With CBT patients are encouraged to change their distorted ways of thinking and irrational beliefs. If a patient is convinced that ``neighbors are spying on him'', the therapist may try to convince the patient that this is not true. Our suggestion, instead, is to break the link between the idea and the negative content. In other words: ``even if neighbors are spying on you, why is that a bad thing?''

Framed in these terms, our therapeutic proposal might look deceptively simple. The major difficulty we can expect to encounter is due to the fact that the number of associated features might be very large (hundreds). This can be clarified with an example. Let us suppose to have 100 black socks (corresponding to negative features), that we want to turn into white socks (corresponding to positive features). Let us also suppose that, at regular intervals, a random subset of socks is put into a washing machine (corresponding to the associative mechanism of the mind). If the white socks in the washing machine are much fewer than the black socks (as we can expect at the beginning of the treatment), they will take the color released from the black socks and turn black again. Only when the white socks become the majority, is the ``washing machine effect'' expected to help the therapy, but this may take a long time. 

\section{Conclusions}

The objective of this work was to merge the psychodynamic theory based on defense mechanisms and the theory of dissociation as a response to trauma, to produce a new interpretation of borderline and narcissistic personality disorders. The outcome is a theoretical framework that allows to highlight the features which are shared between the two disorders and the differences between them. Future work will be aimed to develop the model proposed and draw from it further insights for therapy. 


\bibliographystyle{apalike}
\bibliography{lpdtraumas} 

\end{document}